# Levelized Cost of Energy Calculation for Energy Storage Systems

**H. LOTFI, A. MAJZOOBI, A. KHODAEI**  **S. BAHRAMIRAD, E.A. PAASO**
**University of Denver**  **ComEd**
**USA**  **USA**

**SUMMARY**

The levelized cost of energy (LCOE) presents the energy-normalized cost of a generation asset by considering all associated costs (investment and operation) and total generated energy over its life cycle. As LCOE is a levelized value, it provides a quick and easy measure to compare different energy resource technologies with different characteristics. The LCOE calculation for large-scale power plants and distributed generations (DGs) is extensively studied and can be found in the literature. The discussions on the LCOE calculation for energy storage systems, however, is limited. Although still relatively expensive compared to generation technologies, energy storage is gaining significant attention and has been deployed extensively during the past few years, conceivably due to its many benefits such as load shifting, energy arbitrage, and renewable coordination. Therefore, LCOE calculation of energy storage systems plays an important role in economic evaluation of power systems. This paper proposes a method for calculating the LCOE of energy storage, and further provides the sensitivity analysis with respect to changes in capacity, electricity market prices, and efficiency.

Hossein.Lotfi@du.edu

# 1. INTRODUCTION

The levelized cost of energy (LCOE) is defined as the net present value of the entire cost of electricity generated over the lifetime of a generation asset divided by the total generated energy. In other words, the sum of investment costs, production cost, as well as the operation and maintenance (O&M) costs is calculated and divided by the total energy produced over the lifetime of the asset. The LCOE can be used to efficiently determine if a generation unit is economically viable to be installed and to further investigate if the deployed technology cost can break-even over the lifetime of the project.

Besides providing a simple measure that covers all the costs and produced energy values, there are many benefits associated with the LCOE. The LCOE can be used to efficiently compare various generation technologies with unequal lifetimes, capacities, capital costs, fuel costs, and efficiencies. For example, comparison between the LCOE values of wind and solar technologies in a specific location can determine which is more viable from an economical perspective. It should be considered, however, that a particular generation technology may have different LCOE values in different geographic locations due to weather conditions, the cost of land, availability of construction requirements, and labor costs. The LCOE calculations are valid for both large-scale installations and small-scale distributed generation (DG) installations. Figure 1 shows and compares the LCOE values, including minimum, maximum, and average, for various DG technologies. The other remarkable benefit of the LCOE is that it enables cost comparison of the generation technology with the price of electricity grid at the point of connection to the grid. This comparison determines the economic viability of a generation technology to be deployed at a specific point in the electricity grid while further displaying the grid parity. Grid parity is defined as the point at which a DG can produce electricity for customers with the price equal to or less than utility rates [1], thus suggesting more reliance on local generation resources to supply local loads rather than purchasing power from the utility grid.

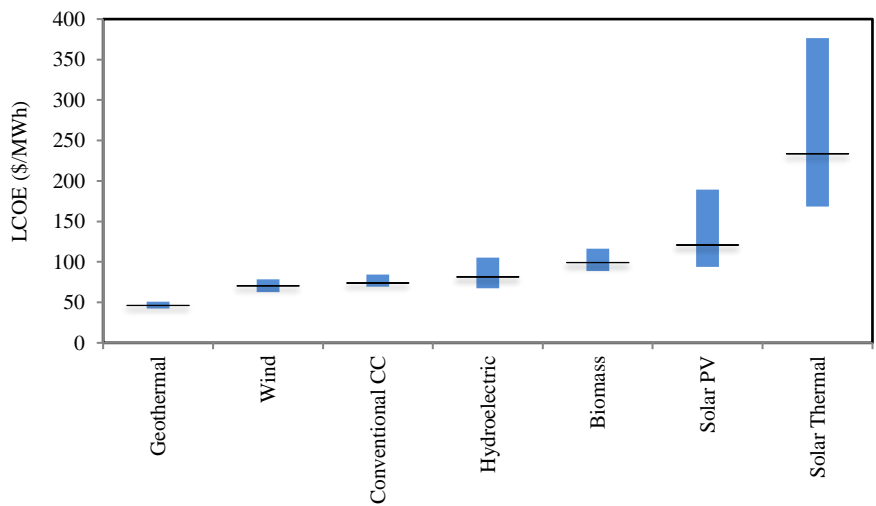

Figure 1. Minimum, average, and maximum LCOE values of various DG technologies

The LCOE calculation of large-scale generation units and small-scale DGs is extensively discussed in the literature. However, the LCOE calculation for energy storage systems, as one of the core constituents of the microgrids and one of the attractive technologies to be deployed on the customer side, is not discussed in detail and is rather limited. Energy storage can be used to store energy in off-peak hours when the electricity price is low and to supply the loads in peak hours when the electricity price is high, hence enabling energy arbitrage and resulting in economic benefits for the owners. Energy storage is also deployed to support renewable generation in which the highly volatile and intermittent generation of these resources can be efficiently captured by coordinated charging/discharging of the energy storage. Further, energy storage is of high importance and applicability in microgrids as a viable means to support islanded operations when the supply of power from the utility grid is interrupted. There are further reactive power supports enabled by energy storage that can be used to correct power factor and/or adjust voltage levels. These benefits, however, come at the expense of the high capital cost of the energy storage. It is therefore extremely important to determine its LCOE and further enable an easy and efficient comparison with other technologies from an economic viability perspective. This paper proposes an LCOE calculation method for energy storage.

The rest of the paper is organized as follows: Section 2 explains various factors that should be considered for LCOE calculation, Section 3 proposes a method for LCOE calculations for the energy storage based on an analogy between a DG and the energy storage, Section 4 contains numerical simulations along with the sensitivity analysis with regards to the time of charging and discharging, the electricity market price, and the energy storage efficiency, and Section 5 concludes the paper.

## 2. SIMPLE MODEL OF LCOE CALCULATION

There are various models to calculate LCOE, presented by different organizations as found in [3]-[5]. All methods are basically based on dividing the sum of the net present costs of capital, fuel, and O&M costs associated with producing energy, by the sum of total energy produced in the asset lifetime. A simplified and general method for calculating LCOE of a generation unit is as follows:

$$LCOE = \frac{(I+F)+(C_F+V)}{E} \qquad (1)$$

where $I$ is the investment cost defined as the capital cost, $F$ is the fixed O&M cost, $V$ is the variable O&M cost, $C_F$ is the unit generation cost, and $E$ is the total energy produced in the project lifetime. In the LCOE equation (1), the investment and fixed O&M costs, i.e., $I$ and $F$, are capacity related terms, as they vary with changes in the installed rated power, while $C_F$ and $V$ are energy related terms as they depend on the energy generated by the unit during its lifetime. The net present costs are obtained using discount rates $d$ which appear as coefficients $1/(1+d)^{t-1}$ for each year $t$.

## 3. LCOE CALCULATION FOR THE ENERGY STORAGE

The mentioned cost and energy terms to calculate LCOE can be directly determined for DGs. In particular, the production cost term represents the cost of purchasing fuel and generating electricity, and the energy term represents the total amount of energy that is produced over the unit lifetime based on the purchased and consumed fuel. Figure 2 shows that fuel, such as gas in gas-fired DGs, can be considered as an input while the produced energy, shown as electricity in the figure, can be considered as an output.

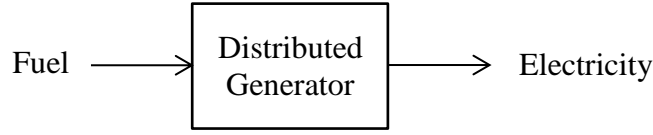

Figure 2. Input and output energies in a DG

For energy storage, however, all of the cost terms cannot be directly determined. For example, for the production cost it is not clear whether the charging cost should be considered, or the discharging benefit, or a combination of both. The same issue exists for the energy as it is not clear whether to consider the charged energy, the discharged energy or net energy. To make these determinations, we use an analogy between energy storage and the discussed DG in terms of inputs and outputs. Figure 3 shows inputs and outputs of the energy storage, depicted similarly to the DG. In DG, the fuel (input) was used to determine the production cost; similarly in the energy storage we can use the charging cost (i.e., the input) to determine the production cost. With the same logic and considering that the DG's output was considered to determine total produced energy, the energy storage total discharged energy will be used to calculate total produced energy by the energy storage.

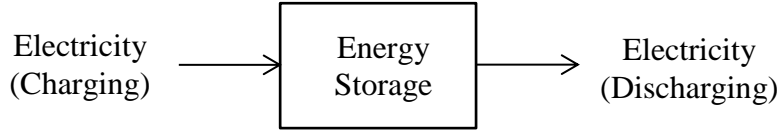

Figure 3. Input and output energies in an energy storage

The formulation for calculating the LCOE of the energy storage can therefore be written as follows:

$$LCOE = \frac{(CC^P P^{\max} + CC^E E^{\max}) + \sum_t \rho E_t^{ch}}{\sum_t E_t^{dch}} \quad (2)$$

where $CC^P$ and $CC^E$ are, respectively, the annualized investment cost for power and energy, $P^{\max}$ is the rated power, $E^{\max}$ is the rated energy, $\rho$ is the utility grid price. $E_t^{ch}$ and $E_t^{dch}$, respectively, denote daily charged and discharged energies, and $t$ is the index of days in a year, i.e., $t=1,\ldots,T$, and $T=365$. This equation can be further simplified based on the following assumptions:
  i. There is a linear relationship between the rated power and the rated energy based on the number of charging hours, $T^{ch}$. This relationship can be written as $E^{\max} = P^{\max} \times T^{ch}$.
  ii. The energy storage is fully charged and discharged each day. Considering the roundtrip efficiency $\eta$, the relationship between daily charged and discharged power can be obtained as $E_t^{dch} = \eta \times E_t^{ch}$.
  iii. The daily charged power is equal to the rated energy, i.e., the storage is fully charged each day, i.e., $E_t^{ch} = E^{\max}$.

Applying these assumptions to the obtained LCOE equation, the simplified LCOE for energy storage can be represented as:

$$LCOE = \frac{(CC^P + CC^E T^{ch}) + \rho T^{ch} T}{\eta T^{ch} T} \quad (3)$$

As the new equation represents, the LCOE of the energy storage only depends on its annualized power and energy investment costs, the charging time, the roundtrip efficiency, and the utility grid price. This LCOE is not dependent on the energy storage size, similar to the LCOE of generation units which is independent of their size. The only external factor is the utility grid price which explicitly shows that the LCOE of the energy storage is highly dependent on its installation location. It is worth mentioning that the proposed model is further capable of considering time-based utility grid prices, which would integrate time dependent value of $\rho$.

## 4. NUMERICAL RESULTS

The data of energy storage used for numerical calculations are represented in Table 1 [6]. The energy storage efficiency and the average electricity price are considered to be 90% and $107.1/MWh, respectively [7]. It is assumed that the energy storage is charged with the maximum capacity in half a day and discharged in the other half. Therefore, the charging/discharging time would be 12 hours. Considering the above data and according to (3), the LCOE of energy storage would be computed as $225.55/MWh.

Table 1 Energy Storage Characteristics

| Allowable Installation Capacity (MW) | Allowable Installation Energy (MWh) | Annualized Investment Cost – Power ($/MW) | Annualized Investment Cost – Energy ($/MWh) |
|---|---|---|---|
| 1 | 12 | 60,000 | 30,000 |

Changes in the LCOE of energy storage with respect to the number of charging hours in a day, the average electricity price, the efficiency, and the ratio of annualized investment cost of energy to power are depicted in Figures 4-7. As can be seen in the figures, increasing the number of charging hours per day would decrease the LCOE as increasing the charging time is associated with a higher capacity factor (Figure 4), while increasing the average electricity price would increase the LCOE linearly (Figure 5).

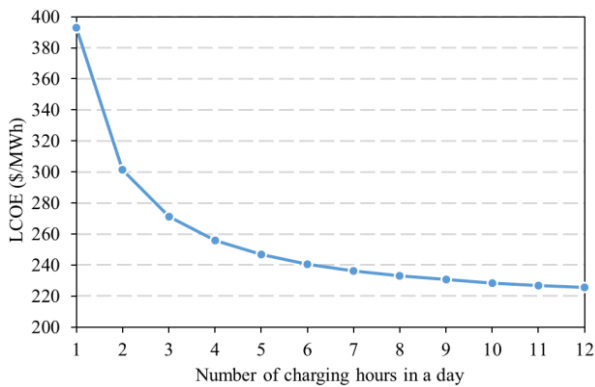

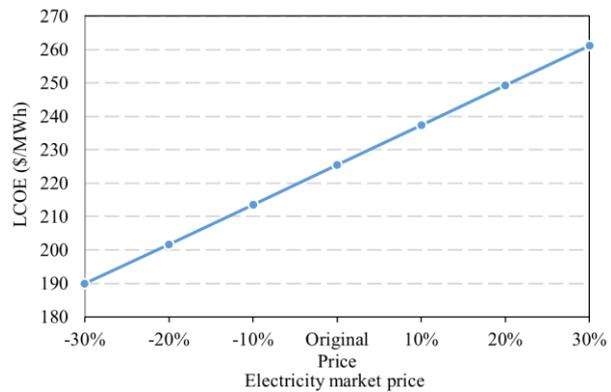

Figure 4. LCOE of energy storage with respect to number of charging hours in a day

Figure 5. LCOE of energy storage with respect to electricity price

It is expected that having a highly efficient energy storage would decrease the total cost, which reduces the LCOE as shown in Figure 6. The results of Figure 7 are based on the initial assumptions ($T^{ch}$=12h, $\rho$=$107.1/MWh, $\eta$=90%), while the annualized investment cost of power was considered the constant amount of $30,000/MW. As this figure shows, the LCOE of energy storage is linearly increased with increasing the aforementioned investment cost ratio.

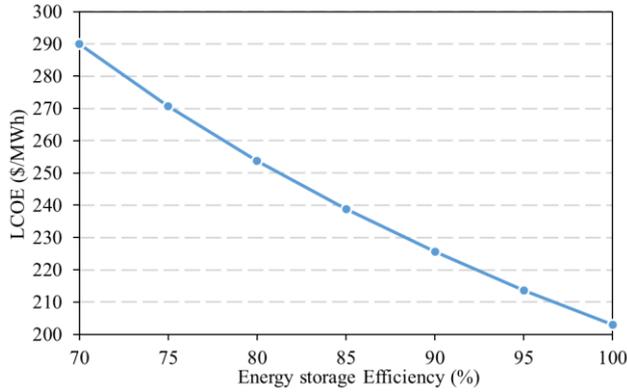

Figure 6. LCOE of energy storage with respect to efficiency

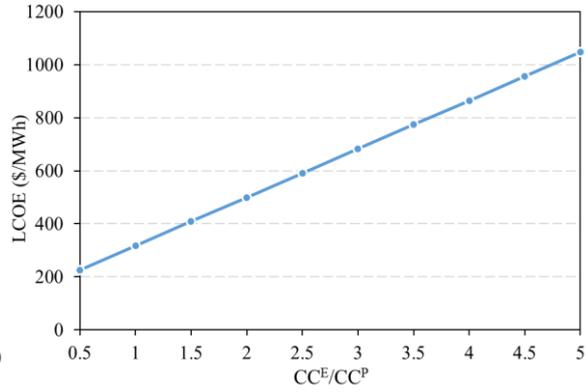

Figure 7. LCOE of energy storage with respect to the ratio of annualized investment cost of energy to power

The range of LCOE values of four common types of energy storage is demonstrated in Figure 8. The amounts of LCOE were calculated based on 12h charging hours in a day and an electricity price of $107.1/MWh. The technical and economical characteristics are borrowed from [8]. The obtained results show that lead-acid and NaS technologies have the lowest LCOE, compared to other commonly-used technologies, i.e., Li-ion and NiCd.

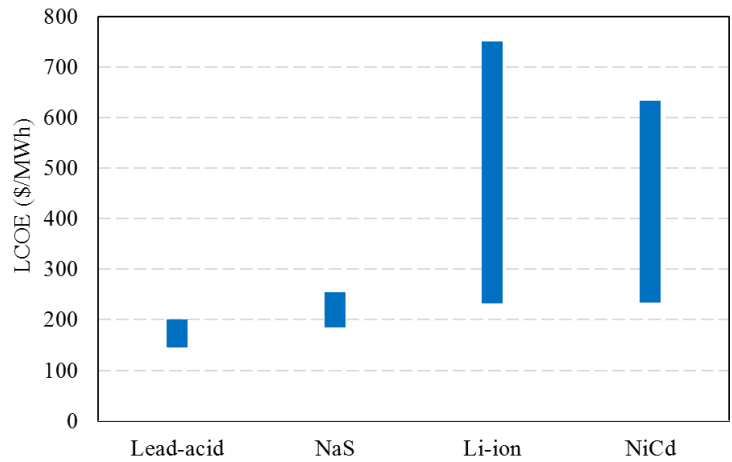

Figure 8. The minimum and maximum LCOE values of various energy storage technologies.

## 5. CONCLUSION

Considering the widespread use of renewable energy sources, the deployment of energy storage in today's power systems, specifically with the significant attention towards microgrids, is of great interest and importance. Energy storage could be charged during off-peak hours when the demand for electricity is low, and discharged during peak hours to improve the system economics. This paper focused on calculating the LCOE of the energy storage to develop a simple and straight-forward comparison method with other types of DERs. The general formulation for LCOE was explained and a method for LCOE calculation of energy storage was proposed based on analogy and according to its specific characteristics. Sensitivity analyses of the obtained LCOE with respect to various parameters were further performed to show the effectiveness of the proposed method. Finally, the range of LCOE for various energy storage technologies was calculated and compared.